\documentclass[aps,pra,showpacs,twocolumn,superscriptaddress]{revtex4}
\usepackage{graphicx,color}
\usepackage{amsmath,amsthm,amsfonts,amssymb,bm}
\usepackage{times}
\usepackage{epsf}
\usepackage[colorlinks={true}]{hyperref}

\usepackage[T1]{fontenc}
\usepackage[utf8]{inputenc}

\hypersetup{citecolor={blue}, filecolor={blue}, linkcolor={blue}, urlcolor={blue}}

\newcommand{\commentold}[1]{}
\DeclareMathSymbol{:}{\mathpunct}{operators}{"3A}

\begin{document}

\title{The Role of the total entropy production in dynamics of open quantum systems in detection of non-Markovianity}
\author{S. Salimi}
\email{shsalimi@uok.ac.ir}
\author{S. Haseli}
\author{A.S. Khorashad}

\affiliation{Department of Physics, University of Kurdistan, P.O.Box 66177-15175 , Sanandaj, Iran}

\date{\today}

\begin{abstract}
In the theory of open quantum systems interaction is a fundamental concepts in the review of the dynamics of open quantum systems. Correlation, both classical and quantum one, is generated   due to interaction between system and environment. Here, we recall the quantity which well known as total entropy production. Appearance of total entropy production is due to the entanglement production between system an environment. In this work, we discuss about the role of the total entropy production  for detecting non-Markovianity. By utilizing the relation between total entropy production and total correlation between subsystems, one can see a temporary decrease of total entropy production is a signature of non-Markovianity.
\end{abstract}


\maketitle

\section{Introduction}
The study of open quantum systems from different perspectives has been the interest of many researchers \cite{bbook,book2,book3}. Information flow between the system and its surroundings is one of the main characteristics of the open quantum systems. Indeed, the flow of information is due to the interaction of the system with its surrounding environment which its direction depends on several factors such as the amount of coupling between system and environment. From the standpoint of the memory effects, the quantum dynamical processes is divided into two categories, namely, Markovian (memoryless) and non-Markovian (with memory) dynamical maps. For Markovian dynamical maps (memoryless process) the coupling between system and environment is weak and  information flow from system to environment continuously. If the coupling between system and environment is strong the memory effects will be apparent and  the future states of the system depend on its past which it is the result of the back flow of information from environment to the system. There is much to learn from the presence of the past in the future, because of this the study of open quantum system in non-Markovian regime, detecting and quantifying that by several distinct criteria , are the major goals of the new researches \cite{rhp,hou,blp,luo,measures}.
One can see the completely positive dynamical map $\Lambda$ is markovian, when it forms a one parameter semi-group with generator $\mathcal{L}$ in Lindblad form \cite{bbook}
 \begin{equation}\label{lindblad}
\mathcal{\widehat{L}} \diamond = -i [H,\diamond] + \sum_{n} \gamma_{n}(2V_{n} \diamond V_{n}^{\dag} -\{V_{n}^{\dag}V_{n}, \diamond\}),
\end{equation}
where $\gamma_{n}\geq0$ for every $n$, $H$ is Hermitian operator and $V_{n}$'s called Lindblad operators acting on the system's Hilbert space. In the case of time dependent, $ \mathcal{\widehat{L}}_{t}$ is referred to time dependent Markovian evolution when $\gamma_{n}(t)\geq0$ for every $t$ and $n$, note that here $ \mathcal{\widehat{L}}_{t}$ does not lead to one parameter semi-group of dynamical maps. Due to this feature violation of semi-group property is not sufficient for dynamical map to be non-Markovian, extra property well known as divisibility, must be violated \cite{piilo}. Divisibility directly corresponds to Markovian dynamics and non-Markovianity can be quantified as the degree of deviation from divisibility of dynamical maps, based on this feature, Rivas et al.  proposed a measure well known as RHP \cite{rhp}. Another phenomena which is used to quantify non-Markovianity is the reduction of distinguishability of quantum states under completely positive trace preserving maps, which is associated to a loss of information about the quantum system. Utilizing this fact, Breuer, Laine and Piilo defined a measure which is called BLP \cite{blp}. A series of measures have been introduced in the light of the behaviour of correlations, both classical and quantum one, under Markovian and non-Markovian dynamical maps \cite{luo,fanchini,rhp,sahar}.
For example, one can refers to the criteria has been proposed by Rivas et al. which is  based on the entanglement between a system and an isolated ancilla \cite{rhp}. Fanchini et al. By using accessible information give an interesting interpretation for this criteria \cite{fanchini}. In the same manner Luo, Fu, and Song, use mutual information to detect non-Markovianity \cite{luo}, their measure have an information interpretation in the context of information loss, which is introduced by Haseli et al. \cite{haseli}. As we know creating correlation, both classical or quantum one, is one result of interaction between system and environment, which is caused disorders in total system. During the interaction total entropy changes and amount of this change is always positive and it is called total entropy production $(TEP)$ as a measure of disorders \cite{tep,Schumacher}. Appearance of $TEP$ is due to the entanglement production between system an environment. In this work according to decoherence model, where an isolated
system $\mathcal{S}$ is coupled to a measurement apparatus $\mathcal{A}$, which in turn interacts with an environment $\mathcal{E}$, we want to utilize  $TEP$ quantity in orther to introduce a measure for non-Markovianity. First, we obtain an lower bound for total correlation changes $\Delta I(\mathcal{S}:\mathcal{A})$, interestingly this lower bound in associated with $TEP$ by an inequality. One can see, if total entropy production decrease monotonically increasing in total correlation is inevitable which is a signature of non-Markovianity.
This paper is organized as follo. In Sec. \ref{II} we introduce
the definitions of the entropy exchange. In section \ref{III}, by making use of the connection between total entropy production and quantum mutual information, we introduce
the definition of the related non-Markovianity measure. In section \ref{IV}, some examples in dynamical model are provided in order to examine the measure. Finally in section \ref{V}, we summarize our discussion and results.
\section{Entropy Exchange}\label{II}
Entropy exchange was introduced by Schumacher \cite{Schumacher} and LIoyd \cite{LIoyd}. The entropy exchange of quantum operation $\Lambda$, with input $\rho$ is defined to be
\begin{equation}\label{entropy exchange}
S_{e}(\rho,\Lambda)=S(\widetilde{\Lambda}[\rho])\equiv S(\rho^{\mathcal{\widetilde{E}}}),
\end{equation}
where $\rho^{\mathcal{\widetilde{E}}}$ is the state of the environment after the operation $\Lambda$. Note that the state of the environment initially is assumed pure. If $\Lambda[\rho]=\sum_{i}F_{i}\rho F_{i}$ then an appropriate form for entropy exchange by introducing a matrix $W$ with following elements

\begin{equation}\label{matrix elements}
  W_{ij}=\frac{Tr(F_{i}\rho F_{j})}{Tr(\Lambda[\rho])},
\end{equation}
can be written as
\begin{equation}\label{app form}
  S_{e}(\rho,\Lambda)=S(W)=-Tr(W\log_{2}W).
\end{equation}
\section{Entropy production}\label{III}
In this section, we introduce a new method of quantifying
non-Markovianity through the total change in the entropy of the total system $\mathcal{\mathcal{SA}}$ and environment$\mathcal{\mathcal{E}}$. Our method is established on the decoherence program, where a quantum system $\mathcal{\mathcal{S}}$ is cou-
pled to a measurement apparatus $\mathcal{A}$, which in turn directly in-
teracts with an environment $\mathcal{E}$. One can consider a quantum
system $\mathcal{S}$ that is initially correlated with the apparatus $\mathcal{A}$ and the state $\mathcal{\mathcal{SA}}$ is in product state with environment $\mathcal{\mathcal{E}}$. Initially, the state $\mathcal{SA}$, total state $\mathcal{SAE}$ and the state of the environment $\mathcal{E}$ is assumed pure. The environment only affects the state of the apparatus $\mathcal{A}$. As
a result of the interaction, there emerges an amount of correlation among the individual parts of the closed tripartite system
$\mathcal{SAE}$, and thus the environment $\mathcal{E}$ acquires information about
the system $\mathcal{S}$ by means of the interaction with the apparatus
$\mathcal{A}$. This setting is graphically sketched in Fig. 1, where the
system $\mathcal{S}$ evolves trivially while the apparatus $\mathcal{ A}$ is in a direct
unitary interaction with the environment $E$ . The final state of
the composite bipartite system $\mathcal{ SA}$ is given by
\begin{equation}\label{evolution}
  \rho^{\mathcal{S\widetilde{A}}}=Tr_{\mathcal{E}}[(I\otimes U^{\mathcal{AE}})\rho^{\mathcal{SAE}}(I\otimes U^{\mathcal{AE}})^{\dagger}]=[I\otimes\Lambda]\rho^{\mathcal{AS}},
\end{equation}
$\Lambda$ is the general completely positive trace preserving map. Let us now describe our strategy to derive our witness to detect non-Markovianity. For a bipartite system $\rho^{\mathcal{XY}}$ quantum mutual information is defined as
\begin{equation}\label{mutual}
  I(\rho^{\mathcal{XY}})=D_{\mathcal{Y}}(\rho^{\mathcal{XY}})+J_{\mathcal{Y}}(\rho^{\mathcal{XY}}),
\end{equation}
here $J_{\mathcal{Y}}(\rho^{\mathcal{XY}})=S(\rho^{\mathcal{X}})-min_{\{\Pi_{i}^{\mathcal{Y}}\}}\sum_{i}p_{i}S(\rho^{\mathcal{X}|\mathcal{Y}})$ is the classical correlation between $\mathcal{X}$ and $\mathcal{Y}$ if an observer doing measurement on $\mathcal{Y}$ and $D_{\mathcal{Y}}(\rho^{\mathcal{XY}})=min_{\{\Pi_{i}^{\mathcal{Y}}\}}\sum_{i}p_{i}S(\rho^{\mathcal{X}|\mathcal{Y}})-S(\rho^{\mathcal{X|Y}})$ denotes quantum discord with respect to measurement on $\mathcal{Y}$, where $S(\rho)=-Tr(\rho \log_{2}\rho)$ and $S(\rho^{\mathcal{X|Y}})=S(\rho^{\mathcal{XY}})-S(\rho^{\mathcal{Y}})$ is the von Neumann and  conditional von Neumann entropy respectively. Using the definition  of the quantum discord in our decoherence program the amount of change in the quantum correlation can be derived as
\begin{align}\label{change}
  \Delta D(\mathcal{S}:\mathcal{A})&=D_{\mathcal{A}}(\rho^{\mathcal{S\widetilde{A}}})-D_{\mathcal{A}}(\rho^{\mathcal{SA}})\\ \nonumber \\\nonumber
                            &= [S(\rho^{\mathcal{S|A}})-S(\rho^{\mathcal{S|\widetilde{A}}})]+ \\ \nonumber \\
                            &+min_{\{\Pi_{i}^{\mathcal{A}}\}}\sum_{i}\widetilde{p}_{i}S(\rho^{\mathcal{S}|\mathcal{\widetilde{A}}})- min_{\{\Pi_{i}^{\mathcal{A}}\}}\sum_{i}p_{i}S(\rho^{\mathcal{S}|\mathcal{A}}). \nonumber
\end{align}
\begin{figure}[t]
\includegraphics[width=0.47\textwidth]{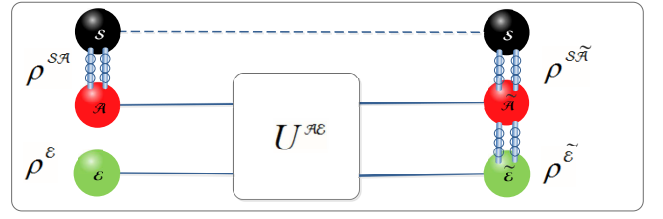}
\caption{We consider an initially pure environment $\mathcal{E}$, and an entangled pure state $\mathcal{SA}$. As the system $\mathcal{S}$ evolves free of any direct interaction, the apparatus is interacting with the environment $\mathcal{E}$.}
\label{fig11}
\end{figure}
We introduce quantity that will
be very useful, namely, total entropy production(TEP) \cite{Schumacher,tep,vedral}. The difference between final total entropy $S_{\widetilde{P}}=S(\rho^{\mathcal{S\widetilde{A}}}) + S(\rho^{\mathcal{\widetilde{E}}})$ and initial total entropy $ S_{P}=S(\rho^{\mathcal{SA}}) + S(\rho^{\mathcal{E}})$ is defined as total entropy production TEP
\begin{figure}[b]
\includegraphics[scale=.7]{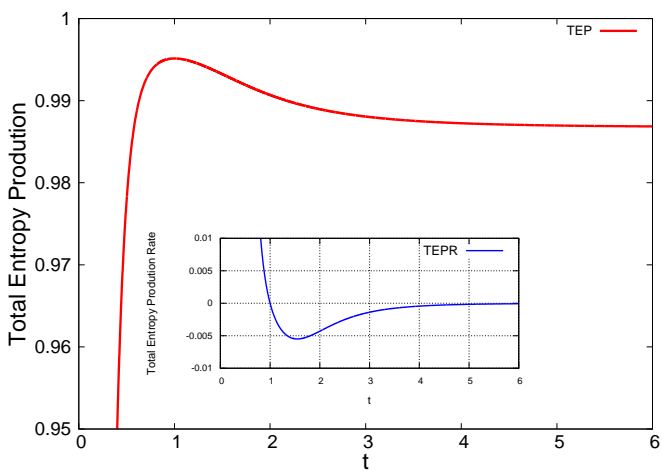}
\caption{(Color online) Total entropy production  as a function of scaled time $\gamma_{0} t$ (red line). The inset displays the behaviour of total entropy production rate $d\Delta S_P /dt$ (blue line) under pure dephasing channel with super Ohmic reservoir spectrum by $s=4$   }
\label{fig1}
\end{figure}

\begin{equation}\label{production}
  \Delta S_{P}=S_{\widetilde{P}}-S_{P}=\Delta S_{\mathcal{SA}}+\Delta S_{\mathcal{E}}.
\end{equation}
Note that (TEP) is always positive. By using the initial and final total entropy and substituting Eq.\ref{production} in to Eq.\ref{change} we have
\begin{align}\label{discord}
  \Delta D(\mathcal{S}:\mathcal{A}) &= -\Delta S_{P}-[J(\rho^{\mathcal{S|\widetilde{A}}})-J(\rho^{\mathcal{S|A}})]- \\ \nonumber \\
                             &-[S(\rho^{\mathcal{AE}})-S(\rho^{\mathcal{\widetilde{A}}})-S(\rho^{\mathcal{\widetilde{E}}})]. \nonumber
\end{align}
According to the definition of mutual information in Eq.\ref{mutual}, one can rewrite Eq.\ref{discord} as
 \begin{align}\label{md}
  \Delta I(\mathcal{S}:\mathcal{A}) &= -\Delta S_{P}-[S(\rho^{\mathcal{AE}})-S(\rho^{\mathcal{\widetilde{A}}})-S(\rho^{\mathcal{\widetilde{E}}})] ,
\end{align}
where $\Delta I(\mathcal{S}:\mathcal{A})=\widetilde{I}-I =\Delta D(\mathcal{S}:\mathcal{A}) + \Delta J(\mathcal{S}:\mathcal{A})$ is the total change in mutual information. Using the fact that $\mathcal{A}$ and $\mathcal{E}$ initially are in product state and subadditivity of the von Neumann entropy and since $\rho^{\mathcal{AE}}$ evolve unitarily we have $ S(\rho^{\mathcal{AE}})\leq S(\rho^{\mathcal{\widetilde{A}}})+ S(\rho^{\mathcal{\widetilde{E}}})$ and Eq.\ref{md} lead to
\begin{equation}\label{final}
  \Delta I(\mathcal{S}:\mathcal{A}) \geq -\Delta S_{P},
\end{equation}
for very small timescales Eq.\ref{final} can be rewritten as
\begin{equation}\label{time}
  \frac{dI}{dt} \geq -\frac{dS_{P}}{dt}.
\end{equation}
Recalling that the LFS measure of non-Markovianity is based on the rate of change of the quantum mutual information shared by the system $\mathcal{S}$ and the apparatus $\mathcal{A}$. In particular, the LFS measure captures the
non-Markovian behaviour through a temporary increase of the
mutual information of the bipartite system $\mathcal{SA}$, from Eq.\ref{time} this satisfy if
\begin{equation}\label{condition}
  \frac{dS_{P}}{dt} < 0.
\end{equation}
It captures the non-Markovian behavior through a temporary decrease of the (TEP) of the tripartite $\mathcal{SAE}$. Mathematically, the measure can be written as
\begin{equation}\label{measure}
  \mathcal{N}_{S_{P}}=\max_{\lbrace\rho_{\mathcal{SA}}\rbrace}\int_{\frac{dS_{P}}{dt} < 0}\frac{dS_{P}}{dt} dt,
\end{equation}
where the maximization is evaluated over all possible pure initial states of the bipartite system $\mathcal{SA}$. Due to the fact that the initial state of tripartite $\mathcal{SAE}$ is pure and remain pure because of unitary evolution, the von Neumann entropy of each subsystem $\mathcal{SA}$ or $\mathcal{E}$ represents the amount of entanglement between them, hence $S_{\widetilde{P}}=2 S(\rho^{\mathcal{\mathcal{\widetilde{E}}}})$, which is equal to twice amount of entanglement between $\mathcal{SA}$ and $\mathcal{E}$ at each instant of the process. Physically, one can say that, if the amount of entanglement between $\mathcal{SA}$ and $\mathcal{E}$ decreases then we confront with the non-Markovian processes. As a result of the Eq. \ref{time}, the process is non-Markovian iff
\begin{equation}\label{fin}
  \frac{dS(\rho^{\mathcal{\widetilde{E}}})}{dt}=\frac{dS(\rho^{\mathcal{S\widetilde{A}}})}{dt}<0.
\end{equation}
where $S(\rho^{\mathcal{\widetilde{E}}})$ is the entropy exchange.
\section{Some Dynamical Examples} \label{IV}
\subsection{Pure dephasing model}
One can employ spin-boson model to construct a general pure dephasing model \cite{bbook,book2,book3,dep}. We consider a two level quantum system as an apparatus $\mathcal{A}$, which linearly interacting with an environment $\mathcal{E}$, so that the total hamiltonian is
 \begin{align}\label{hamiltonian}
  H &= \frac{\omega_{a}}{2}\sigma_{z}+\sum_{i}\omega_{i}b_{i}^{\dagger}b_{i}+\sum_{i}\sigma_{z}(g_{i}b_{i}^{\dagger}+g_{i}^{\ast}b_{i}).
\end{align}
we will focus on the case of initial pure state of the bath. In order to continue the procedure , we have to know the spectral density $J(\omega)$ of the bath. Here we consider the ohmic spectral density as follow
\begin{equation}\label{spectral}
J(\omega)=\frac{\omega^{s}}{\omega_{c}^{s-1}}e ^{\frac{-\omega}{\omega_{c}}}.
\end{equation}
$\omega_{c}$ is cut-off frequency of the spectrum and $s$ is bath parameter. According to this parameter, the bath change from sub-ohmic $s<1$ to Ohmic $s=1$ and super-Ohmic $s>1$ \cite{spectral}. Under these conditions, the reduced dynamic is calculated to give
\begin{align}\label{state}
 \rho(t) &= \begin{pmatrix} \rho_{11} & \rho_{12}\gamma(t)\\ \rho_{21}\gamma(t) & \rho_{22} \end{pmatrix},
\end{align}
where the dephasing parameter $\gamma(t)$ is
\begin{eqnarray}\label{de}
\gamma(t)=\exp[-\int_{0}^{t}\eta(\tau)d\tau],
\end{eqnarray}
where dephasing rate $\eta(\tau)$ is
\begin{eqnarray}\label{f}
\eta(\tau)=\omega_{c}[1+(\omega_{c} \tau)^{2}]^{-\frac{s}{2}}\Gamma(s)\sin  [s \arctan (\omega_{c}\tau)],
\end{eqnarray}
with $\Gamma(s)$ the Euler gamma function. In Fig. \ref{fig1}, we can see the plot of the total entropy production rate $d\Delta S_{p}/dt$ ($TEPR$) (red line), the inset displays the behaviour of $TEP$ (blue line) as a function of scaled time $\lambda t$. in this model. As one can seen, the negative value for toral entropy production rate $TEPR$ is appeared for this model by considering special condition on Ohmicity parameter $s$. Note that in this model dynamic does not admit non-Markovianity for $s<1$.
%\begin{figure}[b]
%\includegraphics[scale=.7]{dep.eps}
%\caption{The entropy diagram of the tripartite system composed of the system $\mathcal{S}$, the apparatus $\mathcal{A}$, and the environment $\mathcal{E}$, before and after the interaction. The amount of information will stay the same inside the area enclosed by thick red curves, i.e., $I=\tilde{I}+\tilde{L}=2S(\rho_{\mathcal{S}})$, where $\tilde{I}$ is the mutual information and $\tilde{L}$ is the quantum loss.}
%\label{fig2}
%\end{figure}
\subsection{Amplitude damping}
Here we consider the apparatus $\mathcal{A}$ as a two-level quantum system  which interacts with zero temperature environment. The dynamics is governed by the following interaction Hamiltonian
\begin{equation}\label{amplitude}
H=\omega_0\sigma_{+}\sigma_{-}+\sum_{k}\omega_k a_k^\dagger a_k + \sum_k (g_k \sigma_{+}a_k +g_k^{*} \sigma_{-}a_k^\dagger)\
\end{equation}
where $\omega_0$ is the transition frequency of the apparatus $\mathcal{A}$ and $\sigma_{\pm}$ denote the raising and lowering operators of the apparatus $\mathcal{A}$. $a_k$ and $a_k^\dagger$ are the annihilation and creation operators of the environment $\mathcal{E}$  , respectively, with the frequencies $\omega_k$.
 The spectral density of the environment has the  Lorentzian form
 \begin{equation}\label{lorantzian}
 J(\omega)= \gamma_0 \lambda^2 / 2\pi[(\omega_0 - \omega)^2 + \lambda^2],
 \end{equation}
 where the $\lambda$ is the spectral width of the coupling.  $\lambda$ is related to the correlation time of the environment $\tau_B$ by $\tau_B\approx 1/\lambda$ and $\gamma_0$ is related to the time scale $\tau_R$ by $\tau_R\approx 1/\gamma_0$, where $\tau_R$ is the time scale that in which the system change. According to these considerations We can define the dynamics of the apparatus via the following master equation as
 \begin{equation}
\frac{\partial}{\partial t}\rho(t)=\gamma(t)\left(\sigma_{-}\rho(t)\sigma_{+}-\frac{1}{2}\{\sigma_{+}\sigma_{-},
\rho(t)\}\right), \label{masterjc}
\end{equation}
where $\gamma(t)$ is the time-dependent decay rate and is given by
\begin{equation}
\gamma(t)=\frac{2\gamma_0\lambda\sinh{(dt/2)}}{d\cosh{(dt/2)}+\lambda\sinh{(dt/2)}},
\end{equation}
where $d=\sqrt{\lambda^2-2\gamma_0\lambda}$.
Thus we can define the dynamics of the $\mathcal{A}$ via the Kraus representation as
\begin{equation}\label{l}
\rho(t)=\Lambda_{t} \rho= \sum_{k=1}^{2} K_{i}(t) \rho K_{i}(t)^{\dagger},
\end{equation}
the corresponding Kraus operators is given by
\begin{align}
 K_1(t) &= \begin{pmatrix} 1 & 0\\ 0 & G(t) \end{pmatrix}, &
 K_2(t) &= \begin{pmatrix} 0 & \sqrt{1-\vert G(t)\vert^{2}}\\ 0 & 0 \end{pmatrix},\label{kraus}
\end{align}
where the function $G(t)$ has the following form
\begin{equation}\label{GF}
G(t)=e^{-\lambda t /2} \left[ \cosh(\frac{d t}{2})+\frac{\lambda}{d} \sinh(\frac{d t}{2})\right].
\end{equation}
\begin{figure}[t]
\includegraphics[scale=.7]{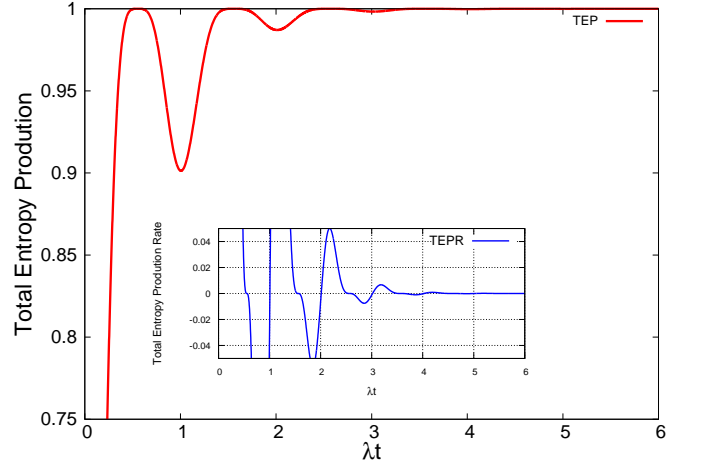}
\caption{(Color online ) Total entropy production as a function of time (red line) and the inset shows the total entropy production rate  $d \Delta S_p / dt$ (blue line) for the exact amplitude damping model with Lorentzian spectral density for reservoir with $\lambda/\gamma_0=0.05$}
\label{fig2}
\end{figure}
As can be seen from Fig.\ref{fig2}, by choosing the value $\lambda/\gamma_0=0.05$, total entropy production rate TEPR in some intervals of time takes the negative value and we have the temporary loss of total entropy production TEP, it shows that in this situation the dynamics is non-Markovian which is consistent with the results from other measures \cite{measures,luo,rhp,blp,fanchini,haseli}.
\subsection{Generalized amplitude damping}
In this case, we consider a two level system as an apparatus $\mathcal{A}$ which interacts with an environment at finite temperature. A generalized amplitude damping channel describes the relaxation due to coupling of system to their surrounding with temperature often much higher than the system \cite{gad}. For single qubit systems it is defined by following kraus operators
\begin{align}\label{ga}
 E_1(t) &= \sqrt{P(t)} \begin{pmatrix} 1 & 0\\ 0 & \sqrt{q(t)} \end{pmatrix}, \nonumber \\
 E_2(t) &= \sqrt{P(t)} \begin{pmatrix} 0 & \sqrt{1-q(t)}\\ 0 & 0 \end{pmatrix}, \nonumber \\
 E_3(t) &= \sqrt{1-P(t)} \begin{pmatrix} \sqrt{q(t)} & 0\\ 0 & 1 \end{pmatrix}, \nonumber \\
 E_4(t) &= \sqrt{1-P(t)} \begin{pmatrix} 0 & 0\\ \sqrt{1-q(t)} & 0 \end{pmatrix},
\end{align}
\begin{figure}[t]
\includegraphics[scale=.7]{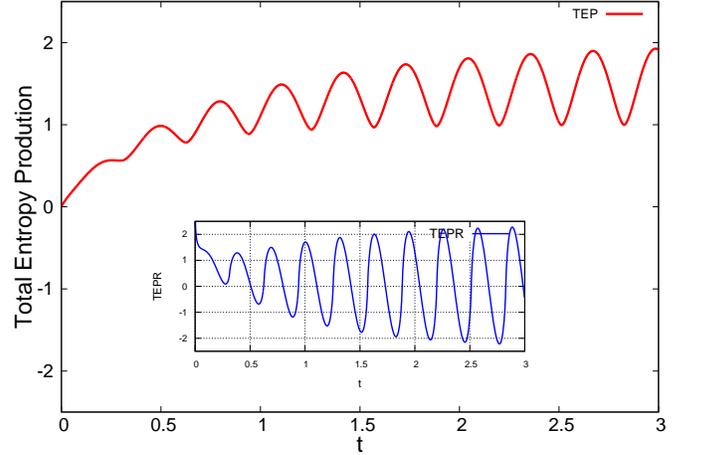}
\caption{(Color online ) Total entropy production  as a function of time (red line) and the inset shows the total entropy production rate $d \Delta S_p / dt$ dynamic (blue line) for the generalized amplitude damping model with $\omega=5$.}
\label{fig3}
\end{figure}
whare for all time $t$ we have $\sum_{i=1}^{4}E_{i}^{\dagger}(t)E_{i}(t)$ and $P(t),q(t)\in [0,1]$. For simplicity we choose the parameters as $P(t)=\cos ^{2} \omega t$ and $q(t)=e^{-t}$, where $\omega$ is a real number. In Fig.\ref{fig3}, one can see for some different time intervals the TEPR is negative, hence TEP decrease and the dynamics is non-Markovian \cite{haseli}.
\section{Results And Discussion} \label{V}
In this paper, our discussion relied on the
decoherence model, where the system $\mathcal{S}$ was coupled to the apparatus $\mathcal{A}$ and initially the composite system $\mathcal{SA}$ was correlated and in pure state. The environment $\mathcal{E}$ directly just had an interaction with apparatus $\mathcal{A}$ and the system $\mathcal{S}$ remain intact. We began our strategy with total correlation between system $\mathcal{S}$ and apparatus $\mathcal{A}$ and continued by the difference between final and initial quantum discord between them. Finally, after straightforward calculation we obtained the inequality between the time derivative of mutual information and entropy production which by making use of the Luo measure \cite{luo} we defined the new criteria in order to detect non-Markovianity based on entropy production. Here, we can give a physical interpretation to this criteria. Because of the interaction between environment $\mathcal{E}$ and apparatus $\mathcal{A}$ the decoherence will be occur and we have the entanglement production between $\mathcal{SA}$ and $\mathcal{E}$. If the amount of the produced entanglement between $\mathcal{SA}$ and $\mathcal{E}$ decreases then we confront with the non-Markovian processes.

\textrm{Acknowledgements.—}The authors acknowledge F.F. Fanchini for useful discussions.


\begin{thebibliography}{5}
\bibitem{bbook} H.-P. Breuer and F. Petruccione, \textit{The Theory of Open Quantum Systems} (Oxford University Press, Oxford, 2007)
\bibitem{book2} R. Alicki and K. Lendi, \textit{Quantum Dynamical Semigroups and Applications} (Springer, Berlin, 2007)
\bibitem{book3} \'{A}. Rivas and S. F. Huelga, \textit{Open Quantum Systems, An Intorduction} (Springer, Heidelberg, 2012).
\bibitem{piilo} J. Piilo, S. Maniscalco, K. Harkonen, and K.-A. Suominen, Phys. Rev. Lett. \textbf{100}, 180402 (2008).
\bibitem{rhp} \'{A}. Rivas, S. F. Huelga, M. B. Plenio, Phys. Rev. Lett. \textbf{105}, 050403 (2010).
\bibitem{hou} S. C. Hou, X. X. Yi, S. X. Yu, and C. H. Oh, Phys. Rev. A \textbf{86}, 012101 (2012).
\bibitem{blp} H.-P. Breuer, E.-M. Laine, J. Piilo, Phys. Rev. Lett. \textbf{103}, 210401 (2009).
\bibitem{luo} S. Luo, S. Fu, and H. Song, Phys. Rev. A \textbf{86}, 044101 (2012).
\bibitem{measures} A. K. Rajagopal, A. R. Usha Devi, and R. W. Rendell, Phys. Rev. A \textbf{82}, 042107 (2010); X.-M. Lu, X. Wang, C. P. Sun, Phys. Rev. A \textbf{82}, 042103 (2010); S. Lorenzo, F. Plastina, and M. Paternostro, Phys. Rev. A \textbf{88}, 020102(R) (2013); B. Bylicka, D. Chru\'{s}ci\'{n}ski, S. Maniscalco, Sci. Rep. \textbf{4}, 5720 (2014).
\bibitem{sahar} S. Alipour, A. Mani, A. T. Rezakhani, Phys. Rev. A 85, 052108 (2012)
\bibitem{fanchini} F. F. Fanchini, G. Karpat, B. \c{C}akmak, L. K. Castelano, G. H. Aguilar, O. J. Far\'{\i}as, S. P. Walborn, P. H. Souto Riberio, and M. C. de Oliveira. Phys. Rev. Lett. \textbf{112}, 210402 (2014).
\bibitem{haseli}S. Haseli, G. Karpat, S. Salimi, A.S. Khorashad, F. F. Fanchini, B. Çakmak, G. H. Aguilar, S. P. Walborn, P. H. Souto Ribeiro, Phys. Rev. A. 90, 052118 (2014)
\bibitem{Schumacher}B. W. Schumacher, Phys. Rev. A 54, 2614 (1996)
\bibitem{tep} T. Sagawa, M. Ueda, New J. Phys. 5, 125012 (2013)
\bibitem{vedral} V. Vedral, J. Phys. Conf. Ser. 143 012010 (2009)
\bibitem{LIoyd}S. LIoyd e-print quant-ph/9604015
\bibitem{dep}G. M. Palma, K.-A. Suominen and A. K. Ekert, Proc.
Roy. Soc. Lond. A 452 567 (1996).
\bibitem{spectral} J.A. Leggett, Rev. Mod. Phys. 59, 1 (1987).
\bibitem{gad} J. Liu, X.-M. Lu, and X. Wang, Phys. Rev. A 87, 042103 (2013).


\end{thebibliography}
\end{document}